\documentclass[twocolumn,prb,showpacs]{revtex4}
\usepackage{epsfig}
\usepackage{amsmath}
\usepackage{amssymb}

\usepackage{color}

\begin{document}

\title{Coherent defect-assisted multiphonon intraband carrier
relaxation in
semiconductor quantum dots}
\author{A.N.~Poddubny$^1$ and S.V.~Goupalov$^{1,2}$}
\affiliation{$^1$~A.F.~Ioffe Physico-Technical Institute, Russian
Academy of Sciences, St. Petersburg 194021, Russia\\
$^2$~Department of Physics, University of Utah, Salt Lake City, UT 84112, USA}
\begin{abstract}
A new defect-assisted mechanism of multiphonon intraband carrier relaxation
in semiconductor quantum dots,
where the carrier is found in a coherent superposition of the
initial, final, and defect states, is proposed. It is shown that this mechanism
is capable of explaining the observed trends in temperature dependences
of the intraband relaxation rates for PbSe and CdSe colloidal nanocrystal
quantum dots.
\end{abstract}
\maketitle

\section{Introduction}
Recent experimental studies of PbSe nanocrystals (NCs)
have revealed an unexpectedly fast intraband relaxation
in semiconductor quantum dots (QDs) with energy separation between
adjacent electron (or hole) levels exceeding by far the optical phonon
energies~\onlinecite{harbold,schaller}. A pronounced temperature dependence
of the observed relaxation rates~\onlinecite{schaller} suggests that underlying
mechanism should involve multiphonon transitions while Auger-like relaxation
processes~\onlinecite{auger} are ruled out due to a nearly symmetric energy spectra in the
conduction and valence bands of PbSe.

The multiphonon transitions between
otherwise orthogonal quantized electronic states in a QD may occur
due to nonadiabaticity of the electron-phonon
system~\onlinecite{hr,ridley,apy,svg05}. A model study performed in conjunction with
the experiments on PbSe NCs revealed that, in order for this mechanism
to be responsible for the experimentally observed temperature dependence
of the relaxation rate, the electron-phonon coupling should be stronger
than one would expect~\onlinecite{svg05}. However, the study of detailed
models of electron-phonon interaction in quantum dots (upon which the
judgement about the coupling strength is based) has always been a weak point
in approaches to energy relaxation~\onlinecite{liara} and Raman~\onlinecite{krauss}
processes. In case of PbSe NCs the situation is further obscured by the fact
that the widely accepted model of their electronic structure~\onlinecite{kang}
which would be natural to pick for an estimate of electron-phonon coupling
strength fails to describe absorption spectra of PbSe NCs~\onlinecite{harbold}
and is therefore not quite reliable. Thus, instead of seeking to either
confirm or rule out this mechanism we will concentrate on examining other
possibilities for the fast energy relaxation.

As it has been proposed for epitaxially grown QDs~\onlinecite{sercel95}, such
possibilities can be provided by a localized state of an impurity
or defect close to the QD surface. In case of the experiment of
Ref.~\onlinecite{harbold} such states can correspond to deep impurities
in the silica host surrounding PbSe NCs occurring in an appreciable proximity
to the NC surface. Otherwise they can be due to surface states.
An electron trapped to the impurity is strongly localized
which assures its strong coupling to local lattice vibrations. One can then consider
a relaxation process involving the interior QD levels as the initial ($2$)
and final ($0$) states, and the state at the impurity as an
intermediate ($1$)
state as shown in Fig.~1. Although processes of this type have been
theoretically investigated~\onlinecite{liara,sercel95,sercel96}, all these studies
considered the relaxation as a two step process, so the relaxation time
$\tau\equiv\tau_{2 \rightarrow 0}=\tau_{2 \rightarrow 1}+\tau_{1 \rightarrow 0}$. But
if $\tau_{1 \rightarrow 0}$ is shorter than the phonon lifetime in state $1$
then the transition $2 \rightarrow 0$ must be treated as a coherent
quantum mechanical process. Such process represents a new mechanism of
phonon-assisted carrier relaxation in QDs and its studying is the scope of
the present work. Similar processes have been considered for capture or generation
of electron-hole pairs by impurities in bulk semiconductors~\onlinecite{apya,inobzor}

Another system where the proposed mechanism can play the key role is provided
by CdSe colloidal nanocrystals. In the recent experiments on such quantum
dots the fast intraband relaxation of electrons has been observed in
the absence of the holes~\onlinecite{sionist}. The electrons were injected into
the quantum dots instead of being optically generated, and the pump-probe
measurements were subsequently carried out~\onlinecite{sionist}. These experiments
showed that intraband relaxation times vary significantly with surface
modification suggesting involvement of a surface state such as the empty
$s$~-~state of the surface Cd$^{2+}$ atoms~\onlinecite{sionist}. A faster relaxation
rate was observed for lower temperatures~\onlinecite{sionist} though more thorough
studies of the temperature dependences of the relaxation rate in CdSe colloidal
nanocrystals are desirable.

It has been noticed~\onlinecite{svg05} that multiphonon relaxation processes
are analogous to tunneling problems in electron transport utilizing
the tunneling hamiltonian.
It turns out that the relaxation mechanism
proposed in this paper is analogous to the problem of electron tunneling
through a potential barrier placed between two leads and containing localized
electronic states coupled to phonons and resonant with the energy of
the tunneling electron. The latter problem was studied by Glazman and
Shekhter~\onlinecite{glazman}. There is, however, a substantial difference
between the two problems due to the descrete energy spectrum for carriers
confined in a QD. Indeed, in the problem of electron tunneling there is a
continuum of both initial and final states in the leads. Thus, this problem
has a characteristic time related to the elastic tunneling of the electron
from the localized state to either of the leads and determining the width
of the resonance. The rate of the phonon-assisted tunneling can be then
calculated using some generalization of the Fermi golden rule with the
$\delta$-function accounting for the energy conservation being integrated
out while summing over the continuum of final electronic states. In our
problem the initial and final electronic states are descrete. However, the
spectrum of the entire system ``electron plus phonons'' is continuous
and the scheme of calculation of the transition rate proposed in
Ref.~\onlinecite{glazman} can still be applied, although some changes should be
introduced. First, the width of the resonance will be determined by the lifetime
of phonons in the intermediate state. Second, the phonon density of
states will enter the expression for the transition rate.
It turns out that these changes can be more naturally incorporated
using a more traditional approach based on calculation of
overlap integrals of eigenfunctions belonging to shifted parabolic
potentials (see e.g.~\onlinecite{sercel96}). Another advantage of this method is that
it allows one to make some qualitative conclusions without having done
actual calculations.

In this paper
the proposed mechanism of carrier relaxation will be studied in detail
for a model QD system while discussions of its relevance to particular
realizations of QDs are left for future studies. The rest of the paper is
organized as follows. In Sec.~\ref{Model} we formulate the problem in a
form most convenient for comparison with the work by Glazman and
Shekhter~\onlinecite{glazman} and with the case of nonadiabaticity-induced
phonon-assisted transitions between intrinsic QD states~\onlinecite{svg05}.
The relaxation rate is expressed through a correlation function which
should be calculated taking into account phonon decay. In Sec.~\ref{GAMMA}
we devise an approximation allowing one to avoid evaluation of this correlation
function. The lifetime of the intermediate state in the relaxation process
is derived in a consistent way. In Sec.~\ref{RATE} results of Sec.~\ref{GAMMA}
are implemented into calculation of the relaxation rate using direct
evaluation of the overlap integrals. Dependences of the relaxation rate on
temperature and electron-phonon coupling strength are discussed in
Sec.~\ref{RESULTS}.

\section{Model}\label{Model}
The Hamiltonian describing multiphonon nonradiative transitions
between three electronic levels with unperturbed energies
$\varepsilon_0$,  $\varepsilon_1$,  $\varepsilon_2$ can be represented as $ H=H_0+V $,
where $H_0$ is the phonon Hamiltonian describing local vibrations in the electronic state with
the energy $\varepsilon_2$,
\[
H_0=\omega_0 b^\dag b \,,
\]
and
\begin{equation}\label{v}\begin{split}
V=\varepsilon_0 a^\dag a+
[\varepsilon_1+ \Delta(b^\dag+b)] c^\dag c
+\varepsilon_2 d^\dag d+\\ (B_{01}a^\dag c+ H.c.)+
(A_{12}c^\dag d+ H.c. )\:.
\end{split}
\end{equation}
Here $a$, $c$, $d$ are the electronic annihilation operators of the
electronic states with the energies $\varepsilon_0$, $\varepsilon_1$, $\varepsilon_2$,
respectively,  and $b$ are the phonon annihilation operators.
We will first neglect
the phonon decay and introduce it later on.
Parameter $\Delta$ determines the shift between the minima
of the vibrational potentials (corresponding to local vibrations
and associated with different electronic states) whose dependence on the
configuration coordinate is sketched in Fig.~1. $A_{12}$ and
$B_{01}$ describe electronic tunneling between the QD states and
impurity level (see Ref.~\onlinecite{sercel95} for details).

We write the electronic wave function as
\begin{equation}\label{ansatz}
\hat\Psi(t)=[ \hat u(t)c^\dag+\hat  v(t)d^\dag+ \hat w(t)a^\dag]|0\rangle \,,
\end{equation}
where  $|0\rangle$ is the electronic vacuum and
$\hat u(t)$, $\hat v(t)$, $\hat w(t)$ are the operators acting on the
phonon subsystem. The wave function~(\ref{ansatz}) satisfies the
Schr\" odinger equation in the interaction representation with respect to the
Hamiltonian $H_0$
\begin{equation}\label{shreq}
i \frac{\partial \hat \Psi(t)}{\partial t}=\hat V(t) \hat \Psi(t)
\end{equation}
with initial conditions
\begin{equation}
\label{initial}
\hat u(t=0)=\hat  w(t=0)=0,\quad \hat  v(t=0)=1.
\end{equation}
Substituting Eqs.~(\ref{v})--(\ref{ansatz}) into Eq.~(\ref{shreq}), we obtain
the following system of equations
\begin{equation}
\label{mainu}
i \frac{\partial\hat  u}{\partial t}=[\varepsilon_1+ \hat{\cal D}(t)]
\hat u(t)+B_{10}\hat w(t)+A_{12}\hat v(t) \:,
\end{equation}
\begin{equation}
\label{mainw} i \frac{\partial \hat w}{\partial t}=\varepsilon_0
\hat w(t)+ B_{01} \hat u(t)\:,
\end{equation}
\begin{equation}
\label{mainv}
i \frac{\partial \hat v}{\partial t}=\varepsilon_2 \hat v(t)+ A_{21}\hat u(t)\:,
\end{equation}
where
\[
\hat{\cal D}(t)=
\Delta (b^\dag \, e^{i\omega_0 t}+
b \, e^{-i\omega_0 t}) \:.
\]
We will follow Refs.~\onlinecite{svg05,glazman} and introduce the auxiliary operator
$\hat u_0(t)$ as the solution of the following Cauchy problem:
\[
i \, \frac{\partial \hat{u}_0}{\partial t}= \hat{{\cal D}}(t) \, \hat{u}_0 \,,
\hspace{2cm} \hat{u}_0(t=0)=1 \,.
\]
In an explicit form it reads
\[\begin{split}
\hat u_0=\exp \Bigl\{
\frac{\Delta}{\omega_0}
[b( e^{-i \omega_0 t}-1)- H.c.]+\\
\left(\frac{\Delta}{\omega_0}\right)^2(i \, \omega_0 t- i \sin \omega_0 t)\Bigr\}
\,.\end{split}
\]
With the help of this operator Eq.~(\ref{mainu})
can be written as
\begin{equation}
\label{u0u}
i \, \frac{\partial}{\partial t} \left( \hat{u}_0^{-1} \, \hat{u} \right)
=\varepsilon_1 \, \hat{u}_0^{-1} \, \hat{u} +
\hat{u}_0^{-1} \, [B_{10}\hat w(t)+A_{12}\hat v(t)]
\,.
\end{equation}

Using Eqs.~(\ref{mainw}),~(\ref{u0u}) the operator $\hat{w}(t)$
can be calculated in the lowest order in the tunneling matrix
elements $B_{01}$, $A_{12}$ as
\begin{multline}
\label{newmainw}
\hat w(t)=-B_{01} \, A_{12} \int\limits_0^t dt_0 \int\limits_0^{t_0} dt_1 e^{i (t_0-t)\varepsilon_0}e^{i (t_1-t_0)\varepsilon_1}\times \\
\, e^{-i \varepsilon_2 t_1}
\hat u_0(t_0) \hat u_0^{-1}(t_1)\:. 
\end{multline}
The probability of the electron transition is given by
\begin{equation}\label{W0}
W_{2 \rightarrow 0}=\frac{1}{\tau_{2\to 0}}=\lim_{t \rightarrow \infty}
\frac{\langle\hat w^\dag(t)\hat w(t)\rangle}{t}\:.
\end{equation}
The average
here is taken over the initial states of the phonon subsystem
at $t=0$. It is assumed that these states are equilibrium
states.
We also have taken into account that, at $t=0$, there is
no interaction-induced re-normalization of the phonon states
since the impurity electronic level is empty.
Substituting Eq.~(\ref{newmainw}) into Eq.~(\ref{W0}) we obtain
\begin{multline}\label{w1}
W_{2 \rightarrow 0}=\lim_{t \rightarrow \infty} \, Z/t \int\limits_0^t dt_0
\int\limits_0^{t_0} dt_1
\int\limits_0^t dt_0' \int\limits_0^{t_0'}  dt_1'
\, e^{-i(t_0'-t_0) \varepsilon_0}\times\\ e^{i (t_1-t_0-t_1'+t_0')\varepsilon_1}
\,
e^{i \varepsilon_2 (t_1'-t_1)}U(t_0,t_1,t_0',t_1')\:,
\end{multline}

where we have introduced the correlation function
\begin{equation}
U(t_0,t_1,t_0',t_1')=\langle \hat u_0 (t_1') \hat u_0^{-1} (t_0')
\hat u_0(t_0) \hat u_0^{-1}(t_1) \rangle\:, \label{bigU}
\end{equation}
and
\[
Z=|B_{01}|^2 \, |A_{12}|^2 \:.
\]
The correlation function~(\ref{bigU}) is analogous to the one which
appeared in the problem studied by Glazman and
Shekhter~\onlinecite{glazman}. It was evaluated in Ref.~\onlinecite{glazman} neglecting phonon
decay. In our case phonon decay should be taken into account as it
determines the width of the resonance corresponding to the
intermediate state in the relaxation problem under study. However, the problem
of evaluating the correlation function~(\ref{bigU}) taking the
phonon decay into account is very difficult. Therefore, in the next
section we devise an approximation which allows one to account for
phonon decay without evaluating correlation function~(\ref{bigU}).
\begin{figure}[h!]
  \centering
    \includegraphics[width=.48\textwidth]{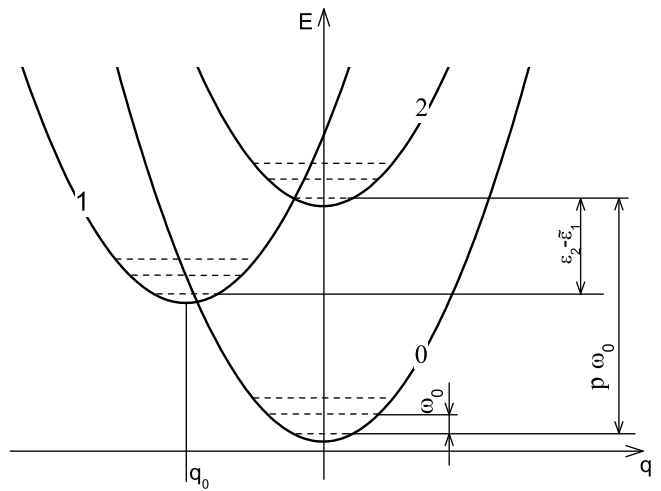}
  \caption{Energy diagram representing adiabatic potentials for the quantum dot states
  (0 and 2) and the impurity level (1).}   \label{curves}
\end{figure}

\section{The width of the resonance}\label{GAMMA}
In order to understand the origin of the width of the
resonance corresponding to the intermediate state in our
relaxation problem we will consider an auxiliary problem
which will help us to apply the concept of quasi-stationary states
to the states of the system ``electron localized at the impurity plus
phonons''. More precisely, let us
set $B_{10}=A_{12}=0$ and consider the electron localized
at the impurity and coupled to phonons.
One can introduce the retarded
Green's function of the electron as
\begin{equation}
G_R(t)=-i \, \langle \langle 0 | \tilde{c}(t) \tilde{c}^{\dag}(0) +
\tilde{c}^{\dag}(0) \tilde{c}(t) |0 \rangle \rangle \, \theta(t) \,,
\label{GR}
\end{equation}
where $\tilde{c}(t)$ is the electron annihilation operator in the
Heisenberg representation and $\langle \langle 0 |...|0 \rangle \rangle$
denotes a quantum mechanical average over the electronic vacuum and a
statistical average over the phonon degrees
of freedom described by the Hamiltonian $H_0$.
As the electron creation operator
$\tilde{c}^{\dag}(0)$ enters the Green's function at $t=0$,
the latter reflects what happens to the electron when it is suddenly
brought to the impurity site from a remote location (where it does
not experience any coupling or entangling to phonons) at $t=0$.
Therefore, this auxiliary problem is essentially equivalent to the
initial relaxation problem. The only difference is that, as
$A_{12}=B_{01}=0$, the auxiliary problem does not describe dynamics of the
electron tunneling to the impurity site.
The Green's function~(\ref{GR}) can be written in terms of the operator
$\hat u_0$~\onlinecite{mahan}
\begin{equation}
\label{GR2}
G_R(t)=-i \, e^{-i \, \varepsilon_1 \, t} \, \langle \hat
u_0(t) \rangle \, \theta(t) \,.
\end{equation}
The spectral function defined as~\onlinecite{mahan}
\begin{equation}
\label{spectral}
A(\omega)=-2  \, \Im \int\limits_{-\infty}^{\infty} dt \, G_R(t) e^{i \omega t}
\end{equation}
consists of a series of $\delta$-functional peaks separated by
the phonon energy
$\omega_0$. Now, if there are processes leading to the phonon decay then
the expression for $\langle \hat u_0(t) \rangle $ becomes~\onlinecite{gslc}
\begin{multline}
\label{u0dec} \langle \hat u_0(t) \rangle= \exp \left\{ - \Delta^2
\, \left[ \frac{\bar{n}_0+1}{(\gamma(\omega_0) + i \omega_0)^2} \:
\left( e^{-(\gamma(\omega_0) + i \omega_0)t} - 1 \right) \right.
\right.
\\
\left. \left. + \frac{\bar{n}_0}{(\gamma(\omega_0) - i
\omega_0)^2} \: \left( e^{-(\gamma(\omega_0) - i \omega_0)t} -1
\right) + \right.
\right.
\\
\left. \left. t \: \frac{\gamma(0) \: (2 \bar{n}_0+1) - i \: \omega_0}
{[\gamma(0)]^2+ \omega_0^2} \right] \right\} \:.
\end{multline}
Here $\bar{n}_0 \equiv \bar{n}(\omega_0)$ stays for the Planckian
factor, $\gamma(\omega)$ is the frequency-dependent phonon decay
rate defined through the imaginary part of the phonon polarization
operator~\onlinecite{gslc}. At zero frequency $\gamma(0)=0$ (which
reflects the fact that the zero phonon absorption linewidth is not
affected by the phonon decay)~\onlinecite{gslc,fn}. Thus, when
(\ref{u0dec}) is substituted into
Eqs.~(\ref{GR2}),~(\ref{spectral}), each peak of the spectral
function, except for the one corresponding to no phonons, becomes
broadened. It is natural to expect that it is this kind of
broadening which causes the finite width of the resonance
associated with the impurity in our relaxation problem.

Assuming
$\gamma(\omega_0)\equiv \gamma\ll\omega_0$, the shape of each peak
of the spectral function can be described analytically. Eq.~(\ref{u0dec})
yields
\begin{multline}
\langle \hat u_0(t)\rangle=\exp\left\{S\left[e^{-\gamma t}[(2 \bar
n_{0}+1)\cos\omega_{0}t-i\sin \omega_0 t)]+\right.\right.\\\left.\left. i \omega_{0}t-(2\bar
n_0+1)\right]\right\}\:, \label{u0dec2}
\end{multline}
where the Huang-Rhys factor $S$ is given by
\[
S=\left(\frac{\Delta}{\omega_0}\right)^2\:.
\]
Substituting Eqs.~(\ref{u0dec2}),~(\ref{GR2}) into
Eq.~(\ref{spectral}) we obtain
\[
A(\omega)=\sum\limits_{l=-\infty}^{\infty}A_l(\omega-\omega_l),\quad
\omega_l=\varepsilon_1-S\omega_0+l\omega_0\:,
\]
\begin{multline}\label{al}
A_l(\omega)=2 e^{l\beta\omega_0/2-(2\bar n_0+1)S} \times \\ \Re\int\limits_0^{\infty}dtI_l[2S\sqrt{\bar n_0(\bar n_0+1)}\exp(-\gamma t)] e^{i\omega t}\:.
\end{multline}
Eq.~(\ref{al}) contains the averaged number of phonons, $\bar{n}_0$.
It is convenient to write out explicitly contributions of the initial
states with given numbers of phonons. To that end we will use
\[
I_l(x)=\left(\frac{x}{2}\right)^l\sum\limits_{k=0}^{\infty}\left(\frac{x}{2}\right)^{2k}\frac1{k!(k+l)!}
\]
and introduce $z=\exp{(-\beta \omega_0)}$ so that $\bar{n}_0=z/(1-z)$.
Substituting this into Eq.~(\ref{al}) we obtain
\begin{multline}
\label{al19}
A_{l}(\omega)=2 e^{-S} \sum_{k=0}^{\infty}\frac{\gamma(2k+l)}{\omega^2+(2k+l)^2
\gamma^2}\times\\\frac{z^k}{(1-z)^{2k+l}}\exp\left(-\frac{2Sz}{1-z}\right) \frac{S^{2k+l}}{k!(k+l)!}\:.
\end{multline}
This expression can be recast using the generating function for associated
Laguerre polynomials~\onlinecite{messiah}
\[
\frac{z^k}{(1-z)^{2k+l}}\exp\left(-\frac{2Sz}{1-z}\right)=
(1-z) z^k \sum\limits_{n=0}^{\infty} z^n L^{2k+l}_n(2S) \,.
\]
As a result, Eq.~(\ref{al19}) can be recast in the form
\begin{equation}
\label{alraz}
A_{l}(\omega)=(1-e^{-\beta\omega_0})\sum\limits_{m=0}^{\infty}A_{lm}(\omega)e^{-m\beta\omega_0 }\:,
\end{equation}
where
\begin{multline}
\label{aldva}
A_{lm}(\omega)=2e^{-S}\times\\\sum\limits_{k=0}^{m}\frac{(2k+l)\gamma}{\omega^2+(2k+l)^2\gamma^2}\frac{S^{2k+l}}{k!(k+l)!}L_{m-k}^{2k+l}(2S)\:.
\end{multline}
Each term in the r.h.s. of Eq.~(\ref{alraz}) corresponds to a transition
from the initial state with $m$ phonons into the impurity state with
$m+l$ phonons and describes this transition's contribution into the
$l$-th peak of the spectral function $A_l(\omega)$.
The latter is contributed
by all processes where exactly $l$ phonons are emitted.
Each of the contributions $A_{lm}(\omega)$ consists of $(m+1)$ Lorentzians.
The Lorentzians with small $k$ in the r.h.s. of Eq.~(\ref{aldva})
can have negative weights while the weight
of the Lorentzian with $k=m$ is always positive. The latter Lorentzian determins
spectral wings of $A_{lm}(\omega)$ and it is natural to associate its width,
$\gamma (2m+l)$, with the linewidth of the transition between the states with
$m$ and $m+l$ phonons. From the theory of spectral
linewidths it is well known~\onlinecite{blp} that the spectral width of
a line corresponding to a transition between two quasistationary states
is given by the sum of their decay rates. As $\gamma$ is the
phonon decay rate and
we have $m$ phonons in the initial state and $(m+l)$ phonons in the
impurity state, we finally arrive at the assignment of the width
$\gamma (m+l)$ to the state of the electron at the impurity entangled to
$(m+l)$ phonons.

\begin{figure*}{t}
  \centering
    \includegraphics[width=.8\textwidth]{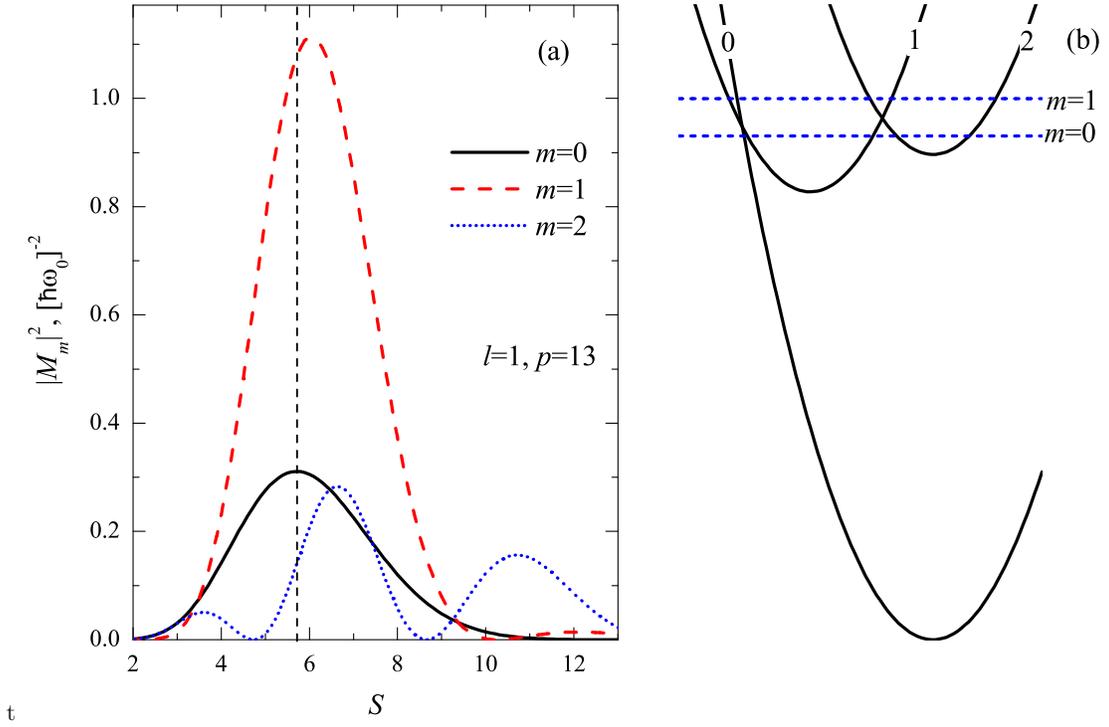}
  \caption{(a)~The dependence of the matrix elements $|M_m|^2$
  on the Huang-Rhys factor, $S$,
  calculated for different values of $m=0,1,2,3$ with $l=1$,
  $p\omega_0=200$ meV, $p=13$, $\gamma=0.2 ps^{-1}$\:.
  The vertical dashed line indicates the maximum of the dependence
  $|M_0|^2(S)$ at $S=5.7$.
  (b)~Adiabatic curves, Eqs.~\eqref{v0}~--~\eqref{v1}, calculated for
  $S=5.7$. The horizontal lines indicate positions of the energy levels
  in the potential $V_2$ with $m=0$ and $m=1$.
  }   \label{m2}  
\end{figure*}
\section{The transition rate} \label{RATE}
We have seen that the lifetime of the intermediate state in the
relaxation problem under study depends
on the number of phonons associated with it. In this case it is
more convenient to consider
the relaxation process in the framework of an approach
based on calculation of overlap integrals between eigenfunctions
belonging to shifted parabolic potentials. These potentials for
our system are schematically depicted in Fig.~\ref{curves}. For
the quantum dot states the potentials are given by~\onlinecite{apy}
\begin{equation}\label{v0}
V_0(q)=\frac{M\omega_0^2q^2}{2}+\varepsilon_0 \,,
\end{equation}
\begin{equation}\label{v2}
V_2(q)=\frac{M\omega_0^2q^2}{2}+\varepsilon_2 \,.
\end{equation}
Here $\omega_0$ is the frequency of the local vibrational mode,
$q$ is the corresponding configurational coordinate, and $M$ is the mass
of the impurity's ion. The adiabatic potential for the intermediate state
is given by~\onlinecite{apy}
\begin{equation}\label{v1}
V_{1}(q)=\frac{M\omega_0^2(q+q_0)^2}{2}+\tilde\varepsilon_{1},
\end{equation}
where
\begin{equation}\label{q0e}
q_0=\sqrt{2S/(M\omega)} \: ,
\end{equation}
\[
\tilde\varepsilon_{1}=\varepsilon_{1} - S \, \omega_0 \,.
\]
Let us introduce the eigenfunctions $\psi^{(2)}_n(q)$, $\psi^{(1)}_n(q)$,
and $\psi^{(0)}_n(q)$ in the adiabatic potentials $V_2(q)$, $V_1(q)$, and
$V_0(q)$, respectively. These are eigenfunctions of the harmonic oscillator
satisfying the following relation
$\psi^{(2)}_n(q)= \psi^{(1)}_n(q-q_0)=\psi^{(0)}_n(q)$. Then the transition
rate $W_{2 \rightarrow 0}$ is given by the Fermi golden rule
\begin{equation}\label{rect}
W_{2 \rightarrow 0}=\frac{2\pi Z}{\omega_0}\langle |M_{2 \rightarrow 0}|^2\rangle \,,
\end{equation}
where
\begin{equation}\label{M2} \langle|M_{2 \rightarrow 0}|^2\rangle=(1-e^{-\beta\omega_0})\sum\limits_{m=0}^{\infty}e^{-\beta m \omega_0}|M_m|^2\:,
\end{equation}
\begin{equation}\label{virtM}
M_m=\sum_{n=0}^{\infty}\frac{\langle \psi_{m}^{(2)} | \psi_{n}^{(1)} \rangle \langle \psi_{n}^{(1)} | \psi_{m+p}^{(2)} \rangle }{(m-n)\omega_0+\varepsilon_2-\varepsilon_1+S\omega_0+i \gamma n}\:.
\end{equation}
The energy $\delta$-function in Eq.~(\ref{rect}) has been integrated
out yielding the phonon density of states $\omega_0^{-1}$ while the value of
$\omega_0$ is adjusted to make the number
$p=(\varepsilon_2-\varepsilon_0)/\omega_0$ integer. We also took into account
that, according to Sec.~\ref{GAMMA}, the width of the intermediate state
is given by the number of phonons in that state times the phonon decay rate.

The overlap integrals entering Eq.~(\ref{virtM})
can be expressed in terms of the generalized Laguerre polynomials~\onlinecite{pbm}
\begin{multline*}
\langle \psi_{m}^{(0)} | \psi_{n}^{(1)}\rangle=(-1)^{(m-n)/2+|m-n|/2}\times\\ e^{-S/2}\sqrt{\frac{\min(m,n)!}{\max(m,n)!}}\sqrt{S}^{|m-n|}L_{\min(m,n)}^{|m-n|}(S)\:.
\end{multline*}
The relaxation is most efficient when the energy of the intermediate state
is in resonance with that of the initial and final states.
We will assume that the resonance condition is always fulfilled and, therefore,
\[
\varepsilon_2-\varepsilon_1+S\omega_0=l\omega_0,\quad l\in \mathbb N\:.
\]
In this case the term with $n=l+m$ becomes dominant in Eq.~\eqref{virtM},
and we obtain
\begin{equation}\label{resM}
M_m\approx\frac{-i}{\gamma(l+m)}\langle \psi_{m}^{(2)} | \psi_{l+m}^{(1)} \rangle \langle \psi_{l+m}^{(1)} | \psi_{m+p}^{(2)} \rangle \:.
\end{equation}


\section{Results and discussion} \label{RESULTS}

\begin{figure}[b!]\label{tt}
  \centering
    \includegraphics[width=.47\textwidth]{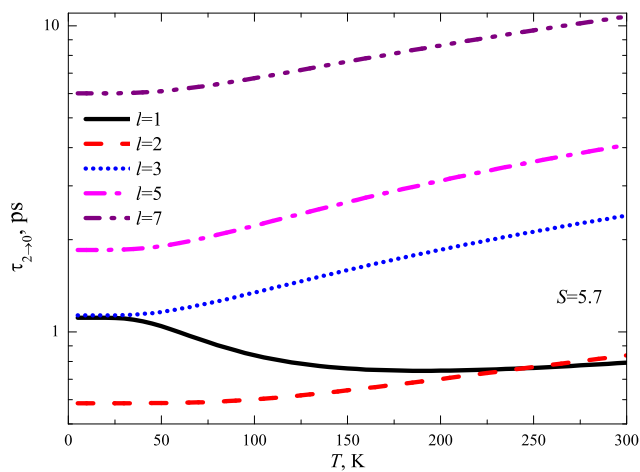}
  \caption{Temperature dependences of the relaxation time
   calculated for different values of the impurity level position
   ($l=1,2,3,5,7$) with $Z=(2.5)^4$ meV$^4$, $S=5.7$. Other parameters
   are the same as in Fig.~\ref{m2}.
  }   \label{tt}
\end{figure}
At zero temperature, only the contribution with $m=0$ survives in
Eq.~(\ref{M2}), and the dependence of the relaxation rate
on the strength of the electron-phonon coupling is governed
by the dependence of $|M_0|^2$ on the Huang-Rhys factor $S$.
This dependence is shown in Fig.~\ref{m2},~(a) (solid curve). Both $S$ and
$\varepsilon_1$ were changed in such a way that the energy
difference $\varepsilon_2-\varepsilon_1+S \omega_0 \equiv l \omega_0$
remained fixed. This is equivalent to a horizontal shift of
the adiabatic curve $1$ in Fig.~\ref{m2},~(b). As one can see from
Fig.~\ref{m2},~(a), the dependence $|M_0|^2(S)$ has a maximum yielding
the optimal value of the Huang-Rhys factor at which
the relaxation is the most efficient. This can be understood using
quasiclassical language. According to the quasiclassical
theory, the overlap integral of eigenfunctions belonging to different
adiabatic potentials is maximum at those energy values where the
adiabatic curves intersect~\onlinecite{apy}. The optimal value of the Huang-Rhys
factor corresponds to the situation where both intersection points
are close to the energy level with $m=0$, as shown in Fig.~\ref{m2},~(b).
This figure also explains why maximum of the curve $|M_1|^2(S)$,
also shown in Fig.~\ref{m2},~(a), is close to the
maximum of $|M_0|^2(S)$.
As temperature is increased, thermal activation leads to increasing
importance of the terms in the r.h.s. of Eq.~(\ref{M2}) with $m>0$.
In the meantime,
relative contribution of the term with $m=0$ to the transition rate decreases
due to the factor of $1-\exp(-\beta\omega_0)$ in Eq.~\eqref{M2}. Therefore,
the character of the temperature dependence of the relaxation rate
is determined by the ratio of $|M_0|$ and $|M_1|$.
One can see from Fig.~\ref{m2},~(a) that, for $l=1$ and Huang-Rhys factor
close to its optimal value, $|M_1|$ exceeds $|M_0|$, and relaxation
time decreases with temperature as shown in Fig.~3 (solid curve).
For $l>1$ intersection points would be closer to the level with $m=0$
than to the level with $m=1$ resulting in $|M_0|>|M_1|$. This
will lead to increasing temperature
dependence of the relaxation time, as shown in Fig.~3.

Therefore, the present mechanism can reproduce both increasing
and decreasing temperature dependences of the relaxation time. Those are
two types of temperature dependences observed
experimentally~\onlinecite{schaller,sionist}. Note that when temperature
sweeps from zero to 300~K, relaxation time in Fig.~3 changes within
one order of magnitude which is consistent with experimental
observations~\onlinecite{schaller,sionist}.

\section{Conclusions}
We have proposed a new defect-assisted mechanism of multiphonon intraband
carrier relaxation in semiconductor QDs,
where the carrier is found in a coherent superposition of the
initial, final, and defect states. Combination of different approaches
has helped us to devise a controlable approximation for
the relaxation rate. We have shown that the proposed mechanism
is capable of reproducing both decreasing, as it was
observed for PbSe NCs~\onlinecite{schaller}, and increasing, as it was
reported for CdSe QDs~\onlinecite{sionist}, temperature dependences of the
relaxation time. For reasonable values of
parameters, the change in the relaxation time with temperature
rise from cryogenic to room temperatures was found within one order
of magnitude, in agreement with experimental observations~\onlinecite{schaller,sionist}.

\acknowledgements{
We are indebted to I.N.~Yassievich for useful discussions and
constant encouragement.
This work was funded by the Russian Foundation for Basic Research
under Grant 07-02-00469-a.
A.N.P. acknowledges support by the Dynasty Foundation--ICFPM.\vspace{4cm}}


\begin{thebibliography}{99}
\bibitem{fn} It is convenient to think of $A(\omega)$ as of an optical
absorption spectrum of a hypothetical system with optical gap at the
frequency of the zero-phonon line.
\end{thebibliography}
\end{document}